\documentclass[11pt,twoside]{article}
\usepackage{baaa-eng}
\usepackage{epsf}
\usepackage{psfig}
\begin{document}
\vskip 1.0cm
\markboth{P. Benaglia}{NT emission from early-type stars}
\pagestyle{myheadings}


\vspace*{0.5cm}
\parindent 0pt{INFORME INVITADO          -- INVITED REVIEW}
\vskip 0.3cm
\title{Non-thermal emission from early-type stars}

\author{Paula Benaglia}
\affil{Instituto Argentino de Radioastronom\'\i a, C.C. 5, Villa
Elisa (1894), y Facultad de Cs. Astron\'omicas y
Geof\'{\i}sicas, UNLP, Paseo del Bosque S/N, (1900) La Plata,
Argentina, pbenaglia@fcaglp.unlp.edu.ar}

\begin{abstract} Massive, early-type stars deposit energy and
momentum in the interstellar medium through dense, supersonic
winds. These objects are one of the most important sources of
ionising radiation and chemical enrichment in the Galaxy. The physical
conditions in the winds give rise to thermal and non-thermal
emission, detectable from radio to gamma rays. In this report the
relevant radiation processes will be described and studies on
particular systems will be presented, discussing the information provided by
multifrequency observations. Future steps aiming at understanding the stellar
wind phenomenon as a whole will be outlined.
\end{abstract}

\begin{resumen} Las estrellas de gran masa entregan energ\'\i a y momento al medio interestelar,
no s\'olo en explosiones de supernova, sino a trav\'es de sus fuertes vientos;
producen radiaci\'on ionizante y son una de las fuentes m\'as importantes de
enriquecimiento qu\'\i mico. En el plasma que forma los vientos tienen lugar
procesos que generan tanto radiaci\'on t\'ermica como no-t\'ermica, detectable
desde el rango de radio hasta rayos gamma. En este informe se describir\'an los
procesos f\'\i sicos generadores de la radiaci\'on, se presentar\'an y analizar\'an
ejemplos de los objetos en estudio, se ver\'a qu\'e informaci\'on proveen las
investigaciones multifrecuencia hacia los mismos, y se discutir\'an los pasos a
seguir en el futuro pr\'oximo, tendientes a completar el entendimiento del
fen\'omeno de los vientos en su conjunto.
\end{resumen}

\section{Introduction}

Early-type stars are characterized by high masses ($\geq$ 8
M$_{\sun}$), large luminosities ($> 10^3$ L$_{\sun}$), and high
superficial temperatures ($ > 10^4$ K). Although their life is
shorter than the one of cooler stars, their influence on the surrounding
interstellar medium (ISM) is enormous, not only because of the ionising power
of their intense UV flux, but also through their strong winds. The winds,
driven by radiation pressure, cause the stars to loose mass during
all their life,  and convey energy and momentum to the ISM. The
winds also contribute to the Galactic chemical enrichment, by
ejecting nuclear matter from the stellar surface. The
stars described correspond to spectral types OB  (O -- B3),
and Wolf-Rayet.

\begin{figure}  
\hbox{
   \psfig{figure=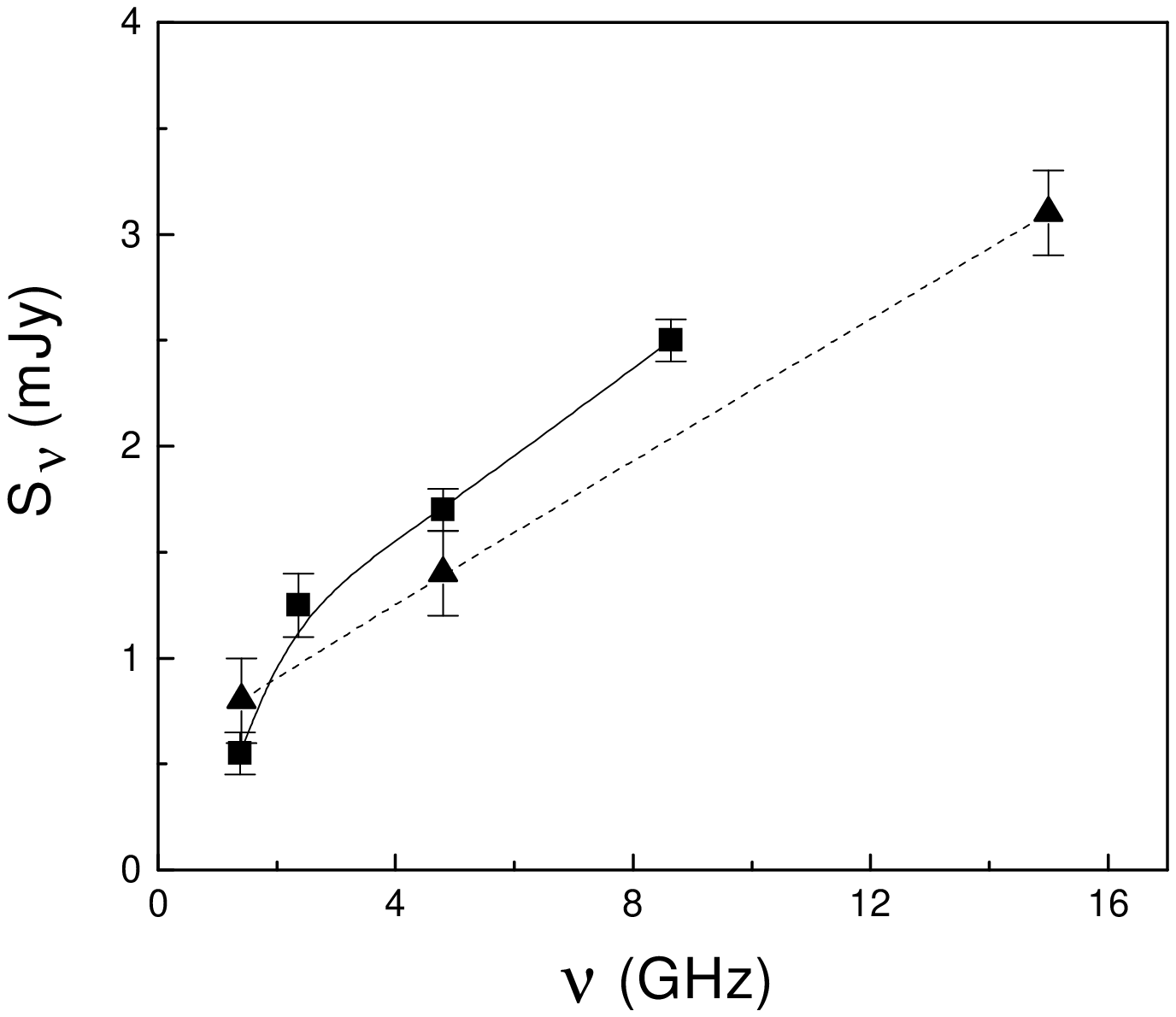,width=6.3cm}
          \hspace*{0.3cm}
   \psfig{figure=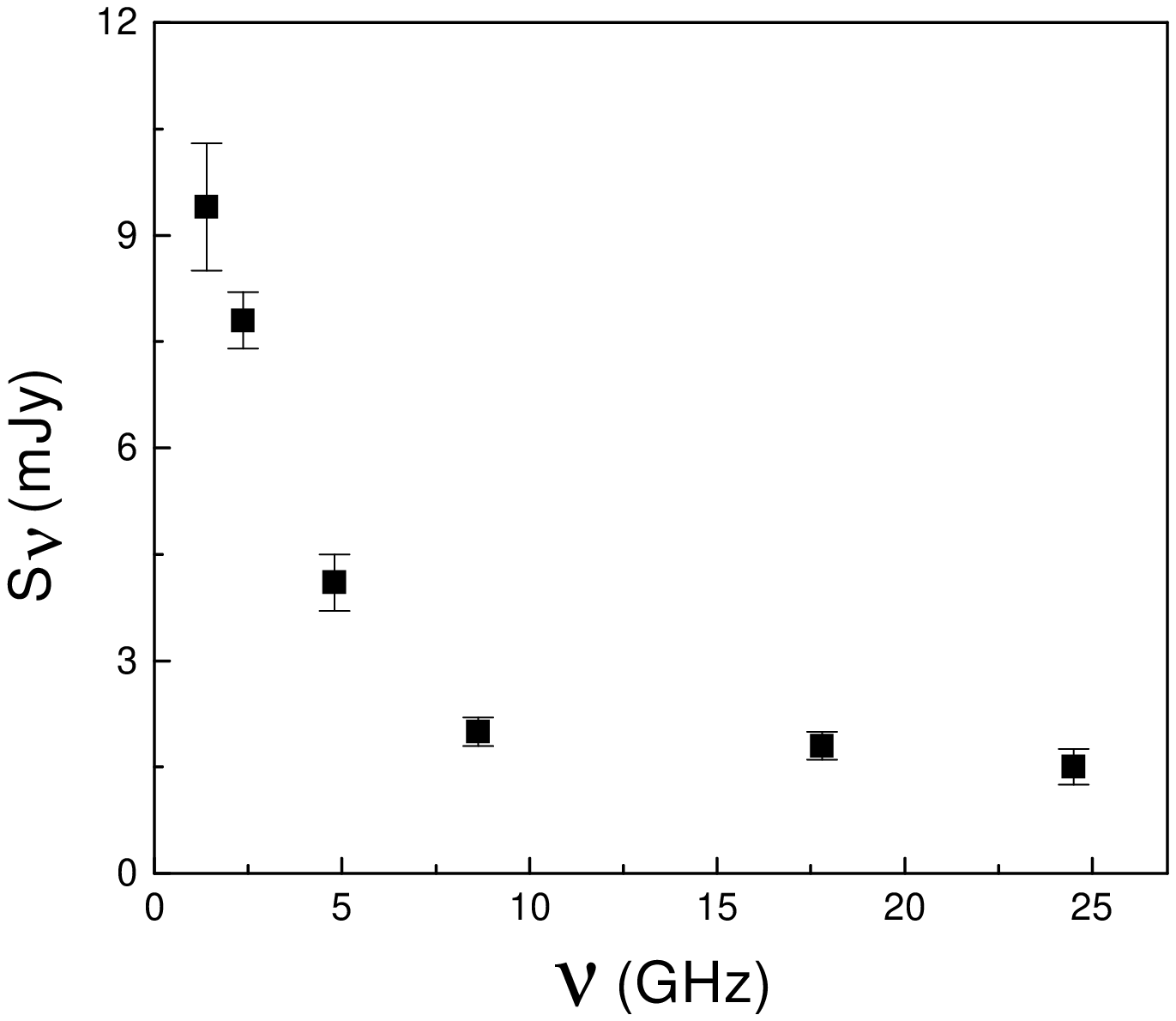,width=6.3cm}
        }
\caption{{\it Left:} Thermal spectra of WR\,40  [WN 8] (squares+solid
line, $\alpha = 0.8$, ATCA data, Chapman et al.  1999), and
$\zeta$Pup [O4 If] (triangles+dotted line, $\alpha=0.6$, VLA data, Bieging
et al. 1989). {\it Right:} Non-thermal spectrum of HD\,93129A
(ATCA data, Benaglia \& Koribalski 2005).}
\end{figure}

Massive, early-type stars (METS) are detectable  at radio
continuum wavelengths, provided high angular resolution ($\sim 1''$), and
sensitivity ($\sim$ mJy) is attainable. The wind itself produces
an excess in continuum radiation, from radio to IR ranges.
This excess has already been measured for many METS (e.g., Bieging et al.
1989, Leitherer et al. 1995) (see Fig. 1-left). The electrons of
the wind plasma emit free-free radiation while decelerate in the
Coulomb field of the ions. This radiation is detectable as  continuum (thermal)
emission. The flux is fitted with a spectral index $\alpha \sim
0.6 - 0.8$ ($S \propto \nu^{\alpha}$), if an ionized,
uniform, isothermal, and stationary gas flow (Wright \& Barlow
1975) is considered. It can be demonstrated that the flux is related to the
stellar mass loss rate $\dot{M}$: the detection of thermal radio
flux offers a straightforward measurement of this stellar
parameter.

However, the radio spectra of some stars present lower -even negative-
spectral indices (e.g. Benaglia et al. 2001b, etc.) Fig.
1-right shows that non-thermal emission is also being radiated from the stellar
winds. What can be learnt from METS non-thermal radio  emission,
mainly by means of radio observations, but complemented with
higher energy studies, is the subject of this
review.

\section {Stars with non-thermal radio emission}

Usually the non-thermal (NT) radio emission presents -at least-
some of the following features: a spectral index value near zero
or negative, radio variability, a radio-derived mass loss rate larger than
the H$\alpha$-derived  rate, and high brightness
temperature.

Information of non-thermal WR, and OB stars can be found in
Benaglia \& Romero (2003) (see their Table 1),  and van Loo (2005)
(see his Table 1.1), respectively. As long as more radio
detections of stellar winds are achieved with better instruments,
the percentage of non-thermal to total emitters has been growing
up to $\sim$ 40\%. Most NT WR stars are not single stars. For OB stars,
there are still a number of NT sources whose binarity status is
unknown.

Under the physical conditions prevailing in the  winds, the
main process responsible of the non-thermal emission detected at radio
waves is synchrotron radiation, i.e., produced by relativistic electrons that
spiral around magnetic field lines. It is assumed that an electron
is accelerated by the first order Fermi mechanism (Bell 1978)
while traversing shocks in the wind.
In single winds the shocks arise from instabilities, triggered, in
turn, by perturbations in particle velocities, regions of
co-rotating interaction,  non-radial pulsations, etc. If the star
has a companion having also a strong wind, and both winds interact with each other, there
will be shocks at the wind collision region (WCR). Shocks can also
be present at the zone where a stellar wind encounters the ISM
(terminal shocks).

\vspace{0.5cm}

\noindent {\bf Single stars}. One of the first wind models for
single stars was introduced by Lucy (1982). It is a phenomenological
model, with outward shocks and adiabatic cooling. White (1985)
proposed the non-thermal emission is synchrotron radiation by relativistic
electrons accelerated in shocks embedded in the wind. The  importance of
inverse Compton cooling was recognized and included by Chen
(1992), and Chen \& White (1994). They could reproduce the
negative radio  spectral index, and concluded that the electrons must be
accelerated in situ at the emitting region. The recent models by
van Loo et al. (see van Loo 2005) consider shocks decreasing in intensity with
stellar distance, narrow emitting layers, and strong shocks that
dominate the emission. They use the latest 1D-hydrodynamical
models to represent the gas flow. The models have trouble to
reproduce some particular cases, for which a yet undetected binary companion
is given as an alternative explanation.

\vspace{0.5cm}

\noindent {\bf Stellar systems with interacting winds}. The
interaction of two supersonic winds will create regions of shocked
gas at high temperatures ($10^7 - 10^8$ K), from which both synchrotron
and free-free emission (f-f) are expected. To detect the non-thermal radiation,  the separation
between binary components must be large enough (periods above some
weeks) in order that thermal electrons from the wind do not bury,
by absorption, the non-thermal emission. Eichler \& Usov (1993)
have physically described the scenario for wide systems. For
radially-flowing winds, a contact discontinuity appears on the
surface where ram pressure of both winds equalize. The position of
this surface can be expressed in terms of the ratio of the
components wind-momentum. The magnetic field in the presence of a
stellar wind is assumed toroidal, and can be computed at the
position of the WCR, if the stellar surface magnetic field values
($B_*$) are known. The authors derive expressions to estimate,
among other quantities, the size of the WCR, the maximum energy
the electrons can gain at the shock, the synchrotron luminosity,
and discuss the production of high energy radiation.

Values of $B_*$ are very difficult to measure; it is possible to estimate an
approximate magnetic field at the WCR by assuming energy
equipartition (Miley 1980).

\section{Current status of radio observations}

>From the observational point of view, the detection of radio fluxes from
stellar winds implies a considerable amount of on-source integration time and the
use of interferometers. The first systematic detection
experiment, carried out by Bieging et al. (1989) using the VLA, yielded 25
detections over 90 northern OB targets. The winds are seen as
point sources for instruments having angular resolution lower than $1''$. Nowadays the observed
OB stars are $\sim$ 150, and about
one third of them have been detected.
In the case of WR stars, circa 90 have been observed, from which
more than 50 were detected at least at one frequency, and about 20
at more than one frequency (Abbott et al. 1986, Chapman et al. 1999, Cappa et al. 2004).

A major breakthrough was achieved in the last decade, when a few
winds could be resolved through VLA, VLBA, and MERLIN data. The
observations corresponded to colliding-wind binary (CWB) systems, and the very WCR could
be imaged.

\vspace{0.2cm}

One of the first systems in which extended sources associated with
stellar winds were detected was {\bf Cyg OB2 No. 5}. This is a
stellar system at 1.8 kpc, formed by a close pair (O7Iaf + Of/WN9)
plus an early B (probable B0 V) (Contreras et al. 1997). The close pair and the B star are
$\sim$ 1700 AU apart. The period of the close pair is $\sim$ 7 days.
The system was observed with the VLA at 5 and 7 GHz (Contreras et
al. 1997, Fig. 2, and references therein), and two sources were
detected: an intense one coincident with the close pair, and a weaker one near
the B star. The stronger source showed radio variability on periods
of 7 yr, switching between a high state with non-thermal emission and
a low-thermal state. The weakest radio source was non-thermal,
and identified as the WCR of the system.

\vspace{0.4cm}

{\bf WR 147} is composed by a WN8(h) star plus an O5-7 star (L\'epine et al.
2001). Its radio image, taken with MERLIN at GHz (Dougherty et al.
1997, Fig. 3, angular resolution: 70 mas) displayed two radio
sources: the northern one has a non-thermal spectral index
($\alpha=-0.5$), and the southern one the characteristic +0.6
thermal index. The northern source has been identified with a WCR
between the components. New MERLIN observations at three epochs
(Watson et al. 2002) revealed that
both the thermal and non-thermal sources vary on timescales of
years.

\vspace{0.2cm}

The binary system of {\bf WR 146} (WC6 + O8) was observed very
recently with the VLA plus one VLBA antenna, attaining an angular
resolution of 30 mas, from  1.4 to 43 GHz. The data were combined
with European VLBI Network (EVN, 9 mas) and 5-GHz MERLIN observations (O'Connor et al.
2005). Fig. 2-left shows two thermal radio sources on the stars,
and a non-thermal one with a bow shock shape corresponding to the WCR.

\vspace{0.2cm}

Maybe the most striking observational radio results were obtained
for {\bf WR 140} (WC7 + O4-5). The period of the system is 7.9 yr
and  the components are separated between 3 and 30 AU. Dougherty
et al. (2005) presented VLBA-8.4 GH data taken during two years,
of a non-thermal ($T \geq 10^7$ K)
 source, moving
along the orbit (Fig. 2-right). The flux depends on the phase. The
radio source was identified with the WCR. The study was
complemented with previous VLA data from 1.4 to 22 GHz. They
derived new orbital parameters, and a new distance to the system 
of 1.85 kpc.

\begin{figure}  
\hbox{
   \psfig{figure=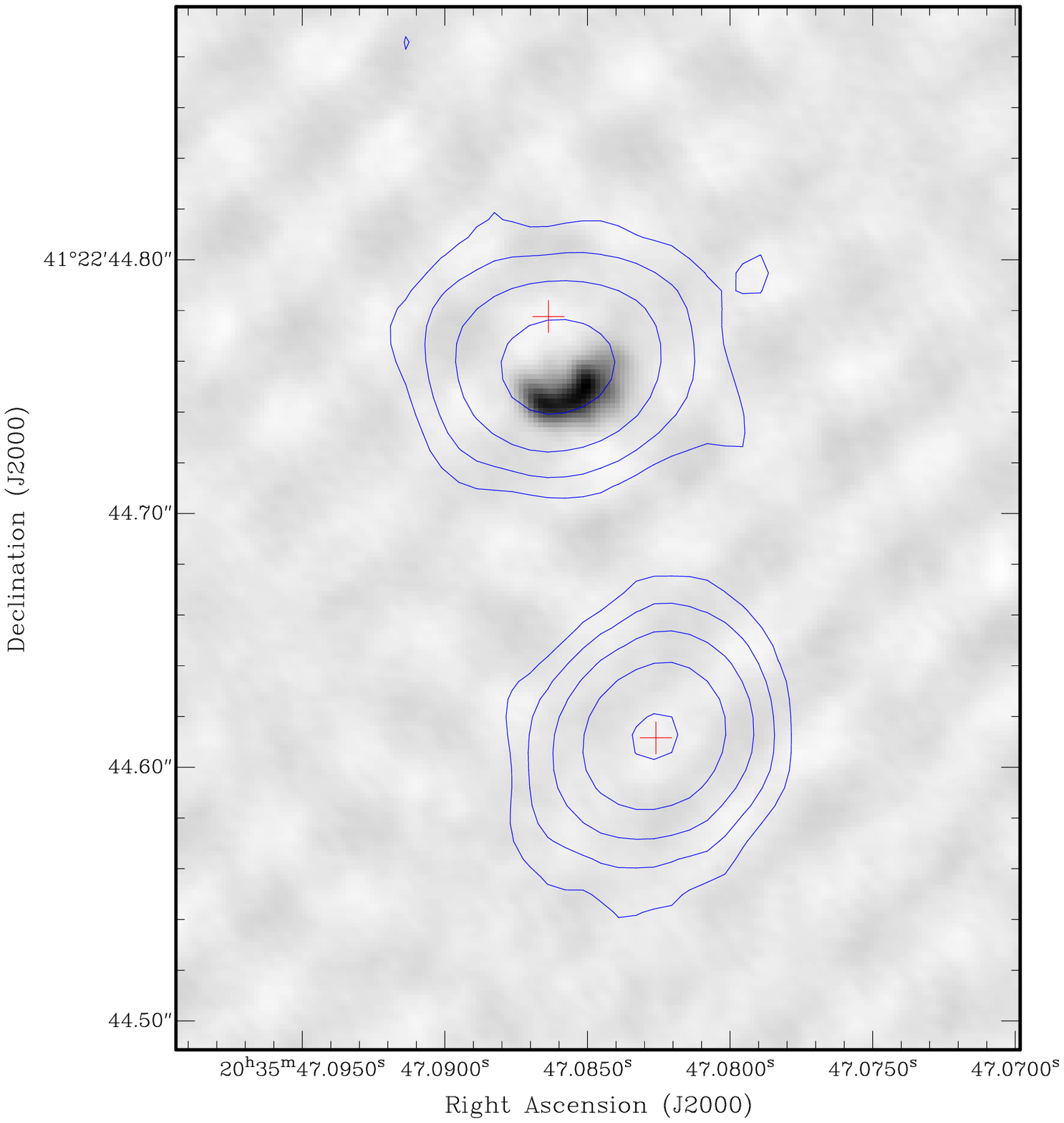,width=5.6cm}
          \hspace*{0.2cm}
   \psfig{figure=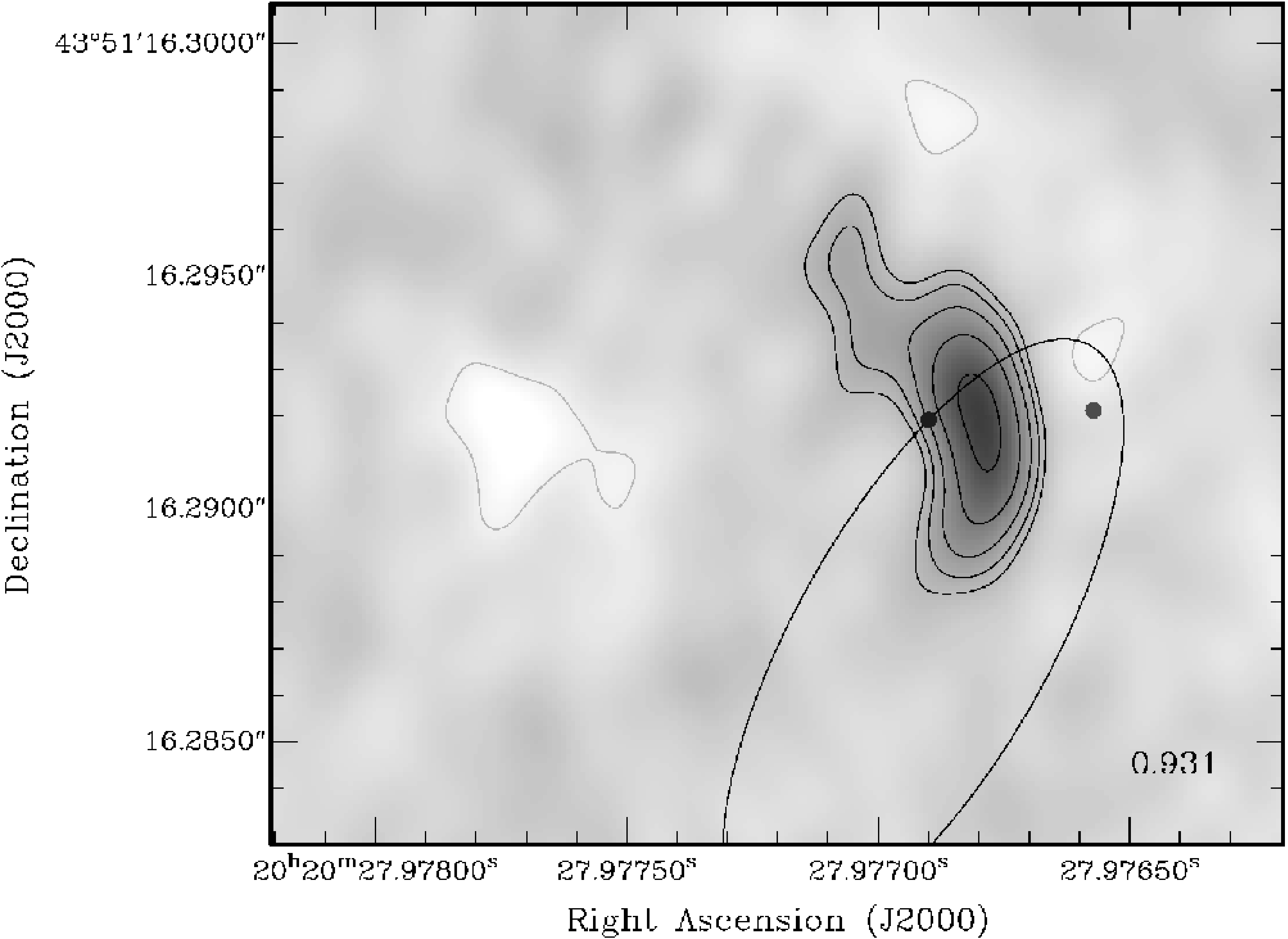,width=7.cm}
        }
\caption{{\it Left:} Overlay of VLA 43 GHz (contours) with EVN
(greyscale) 5-GHz emission on WR 146. The crosses mark the stellar
positions, as derived from HST observations (from  O'Connor et al.
2005). {\it Right:} An 8.4~GHz VLBA observation of WR 140, with
the deduced orbit superimposed. The WR star is to the NW (from
Dougherty et al. 2005).}
\end{figure}

\section{Radio continuum studies}

Radio continuum observations at high angular resolution ($\sim
1''$) allow to derive spectral indices when data at more than
one frequency are available. If the flux can be represented by a  thermal
spectral index, and the basic stellar parameters are known, the
mass loss rate can be derived. In case the index is other than
thermal, the study of the contributions to the radio spectra will
give a hint on the radiation processes involved. An equipartition
magnetic field at the WCR can be estimated, and a value for the
stellar surface magnetic field, extrapolated. Non-thermal sources
could show polarisation, too.

If both angular resolution and sensitivity are high enough, ($< 1''$ and 0.1
mJy, respectively), an {\sl x,y} map of the emitting wind on the plane of the sky can be built.

Anyway, the observations can be used to compare different
emission models, or to create new models in order to interpret the
data. Studies on  high-energy emission from the winds, optical
searches for binary companions, low-resolution radio line
observations over the circumstellar region allow a more complete
description of the stellar winds.

\subsection{Point source approach}

When the stellar wind -or the WCR- is seen as a point source, a simple
model that consist of a thermal source plus a non-thermal source
attenuated by free-free absorption can be used (e.g. Chapman et
al. 1999). The observed flux is expressed as $S_{\rm obs} = S_{\rm
T} + S_{\rm syn} e^{\tau}$, where $\tau$ is the f-f optical depth, at 
frequencies high enough to disregard sychrotron self-absorption
and the Razin-Tsytovitch effects.
The last equation can be solved by means of simultaneous measurements
at different frequencies, and thermal from non-thermal
contributions can be then detached.

This model does not take into account that synchrotron emission
and opacity $\tau$  are probably time variable. Synchrotron
emission will vary if the stellar separation varies as in an
eccentric orbit. The opacity to the synchrotron emission will
change as long as the line of sight moves around the system.

The simplification here is to consider a univalued opacity,
determined through the unique line of sight, and a point synchrotron
source. These models proved to reproduce the radiometry of various
systems (Chapman et al. 1999, Skinner et al. 1998, Benaglia \&
Koribalski 2004, 2005).

\subsection{Extended source approach}

The wind collision regions are excellent places to  study particle
acceleration, because their mass, photon and energy densities are
much higher than in the  SN environment, and the  fundamental
parameters of the wind flow can be derived. A second approach in
the physical description of stellar winds consists of assuming the
WCR as an extended source, and is applicable to the cases where
the spatial distribution of the radio flux can be imaged.

In this approximation the wind region is divided in cells.  A
two-dimension hydrodynamical code is used to characterize the
temperature and density of each cell, which are the input to
compute the absorption and emission coefficients. The equation of
transport for each line of sight is then solved
numerically, suppliying the flux on the plane of the source, for a
given frequency (Dougherty et al. 2003, Pittard et al. 2005).
This process was applied to WR 140 (Dougherty et  al. 2003) and WR
146 (O'Connor et al. 2005) and the agreement between  observations
and predictions show the developments point in the right direction.

\section{High energy non-thermal emission}

Stellar  winds have been recognized as places for particle
acceleration; they are permeated by copious UV flux produced at
the stars. 

When relativistic
electrons interact with UV stellar photons, these photons can be
boosted to higher energies through inverse Compton (IC) scattering processes.

By other hand, additional high energy emission is produced by the same relativistic electrons when they interact 
with the electrostatic field of the nuclei (relativistic Bremsstrahlung).

The same mechanism  efficient to accelerate electrons to
relativistic energies should act on the ions; the interaction between
relativistic nuclei and cold wind nuclei (p-p interaction)
yields to neutral pions that immediately decay generating gamma rays.

At the WCR, the same population of high energy electrons can give
rise both to synchrotron and IC scattering, where $<h \nu_{\rm IC}> =
4/3 \gamma^2 h \nu_*$. The temperatures present at the massive stars we are
dealing with are consistent with seed photon energies of $h
\nu_{\rm *} \sim $ eV which, combined with the electron Lorentz
factors of $10^2$ -- $10^4$ easily attained at the shocks will
give IC photons from keV to MeV energies.

If the electron energy distribution  resulting from a first order
acceleration process can be represented by a power law $N_e(E)
\propto E^{-p}$, then the IC photon distribution will also be a
power law $dN_{\rm ph}(E)/dE \propto E^{-\Gamma}$, where $\Gamma =
(p+1)/2$. In this scenario, the non-thermal spectral index 
$\alpha_{\rm NT} = (p+1)/2$.

The effect of the UV photons generated by the secondary star, closer to
the WCR, will be more important than those from the primary star.
Adiabatic, synchrotron, and IC looses are important 
particle energy looses. The latter two can produce a break at the
electron distribution. The maximum energy gained by an electron at
a certain WCR can be computed if parameters of the involved stars
such as terminal velocities, luminosities, mass loss rates, and
the local magnetic field are known.

The study of the gamma-ray  production at stellar winds is
valuable to find counterparts to unidentified gamma-ray sources,
as the many detected by the EGRET experiment (Hartman et al. 1999, Romero et al. 1999).

\section{Examples of multiwavelength studies}

\subsection{Cyg OB2 No. 5}

This stellar system, described in Sect. 3, is positionally
coincident with the probability contours of the unidentified EGRET
source 3EG J2033+4118. If the gamma-ray source is located at the
stellar distance, it will have a luminosity of $\sim 2.4 \times
10^{35}$ erg s$^{-1}$. The non-thermal  radio emission detected
from the putative wind collision region (Contreras et al. 1997)
reveals the presence of relativistic electrons. The question to
address is whether these electrons and heavier particles through
different interactions could give rise to gamma-rays, and how much
they contribute to the observed EGRET flux. We
evaluated which processes are relevant in the different regions
where shocks can be present in the stellar system (Benaglia et al. 2001a).

Firstly we needed to adopt stellar parameters such as mass loss
rates, wind terminal velocities, and a stellar magnetic field. 
Three regions with shocks have been taken
into account: the WCR between the close pair and the B star, the
stellar winds themselves, and the terminal shock.

It can be demostrated that the
values for the synchrotron and IC luminosities are proportional
($L_{\rm syn} = 840 L_{\rm IC}\,B_{\rm WCR} r_2 / L_2$, ``2''
stands for the secondary; Chen \& White 1994), when they are  produced by the same population of particles. 
The  gamma-ray flux
corresponding to relativistic Bremsstrahlung by electrons involved
in synchrotron processes can be expressed in terms of the
synchrotron flux, as a function of the local electron density and
magnetic field (Benaglia et al. 2001a). At the WCR the values
obtained for the luminosities are $L_{\rm IC} \sim 8 \times
10^{34}$ erg s$^{-1}$, and $L_{\rm rB} < 10^{31}$ erg s$^{-1}$.

In the neighborhoods of Cyg OB2, various CO clouds have been detected
(Dobashi et al. 1996), with masses $\sim 10^3 M_{\sun}$. This observational 
fact allowed to conclude that the
gamma-ray luminosity produced by $\pi^0$ decay by hadrons
``illuminating'' those clouds would sum up to $10^{33}$ erg
s$^{-1}$ (see Aharonian \& Atoyan 1996).
The important contribution in gamma rays at the base  of the wind
is from $\pi_0$ decays, and has been computed by White \& Chen
(1992) as $L_{\pi^0} \sim 5 \times 10^{34}$ erg s$^{-1}$.

The sum of  all luminosities mentioned above can explain about half of the
detected gamma ray flux. We speculate that the remaining flux can be due to the action
of other METS present in the field.

\subsection{WR\,140, WR\,146, and WR\,147}

The (WR+O) binary  systems of WR 140, WR 146, and WR 147 are among the
most studied ones. All have been observed several times and at
different angular resolutions with interferometers. WR 140 is a
8 yr-period binary, while the other two are in very wide orbits,
with probable periods larger than 100 yr. From these stars not
only non-thermal emission has been detected and monitored, but a
map of the radio flux distribution could be built.

WR 140 is  located onto an unidentified EGRET source (3EG
J2022+4317). The estimated EGRET threshold at the positions of WR
146 and WR 147 is high, more than 50\% of the 3EG J2022+4317 
flux value. With
the synchrotron flux and the size of the WCR it is possible to
evaluate the contribution of IC stcattering, relativistic
Bremsstrahlung and $\pi^0$ decay processes producing gamma rays at
the WCR, and compare it with the EGRET results (Benaglia \& Romero 2003).

An improvement has been included in working out the problem, with
respect to the case of Cyg OB2 No. 5, by taking into account the
break in the energy distribution due to synchrotron and IC losses.

The local  magnetic field was the main unknown: we have estimated
an equipartition value for each system. Ultimately, the process
served to calibrate the adopted parameters (see Benaglia \& Romero
2003). It was shown that under reasonable assumptions, {\sl (i)}
the gamma-ray emission from 3EG J2022+4317 could be produced by the
CWB system of WR 140 and {\sl (ii)} the high energy emission from
WR 146 and WR 147 remained below the EGRET detection limit.

As in  the case of Cyg OB2 No. 5, new gamma-ray observations with
better angular resolution and sensitivity are necessary to
fine-tune the assumed physical parameters relevant in this kind of
studies.


\subsection{HD\,92129A}

This  is the only O2 If* cataloged so far (Walborn et al. 2002), a
member of Tr 14 in the Carina region ($\sim 2.5$ kpc, Walborn
1995). HST observations have revealed the presence of an
early-type companion, at 55 mas ($\sim$ 150 AU) (Nelan et al.
2004).

The system was detected at radio continuum from 1.4 to 25 GHz
(Fig. 3-left) over a period of 1 yr (Benaglia \& Koribalski 2005). The radio  spectra displayed
strong non-thermal emission, superposed to the thermal emission
from the hot winds. In order to disentangle both contributions, we
fitted the flux with the expression $S_\nu = S_{\nu}^{\rm T} +
S_{\nu}^{\rm NT} = {\rm C_1} \nu^{0.6} + {\rm C_2}
\nu^{\alpha_{\rm NT}} e^{-\tau_0 \nu^{-2.1}}$, by assuming that 
f-f absorption is modifying the synchrotron emission. On this first 
approximation, the
Razin-Tsytovitch effect and synchrotron self-absorption were
disregarded.

The results allowed not only to characterize  thermal emission and
derive the mass loss rate of the system, but to find the
non-thermal average spectral index which represents the
synchrotron radiation. The assumption of a colliding wind region
size leaded to the estimate of a local magnetic field value of
$\sim 10$ mG.

In the near future, long baseline radio observations will be envisaged, 
to map  the strong NT source that represents the wind-collision region.

\begin{figure}  
\hbox{
      \psfig{figure=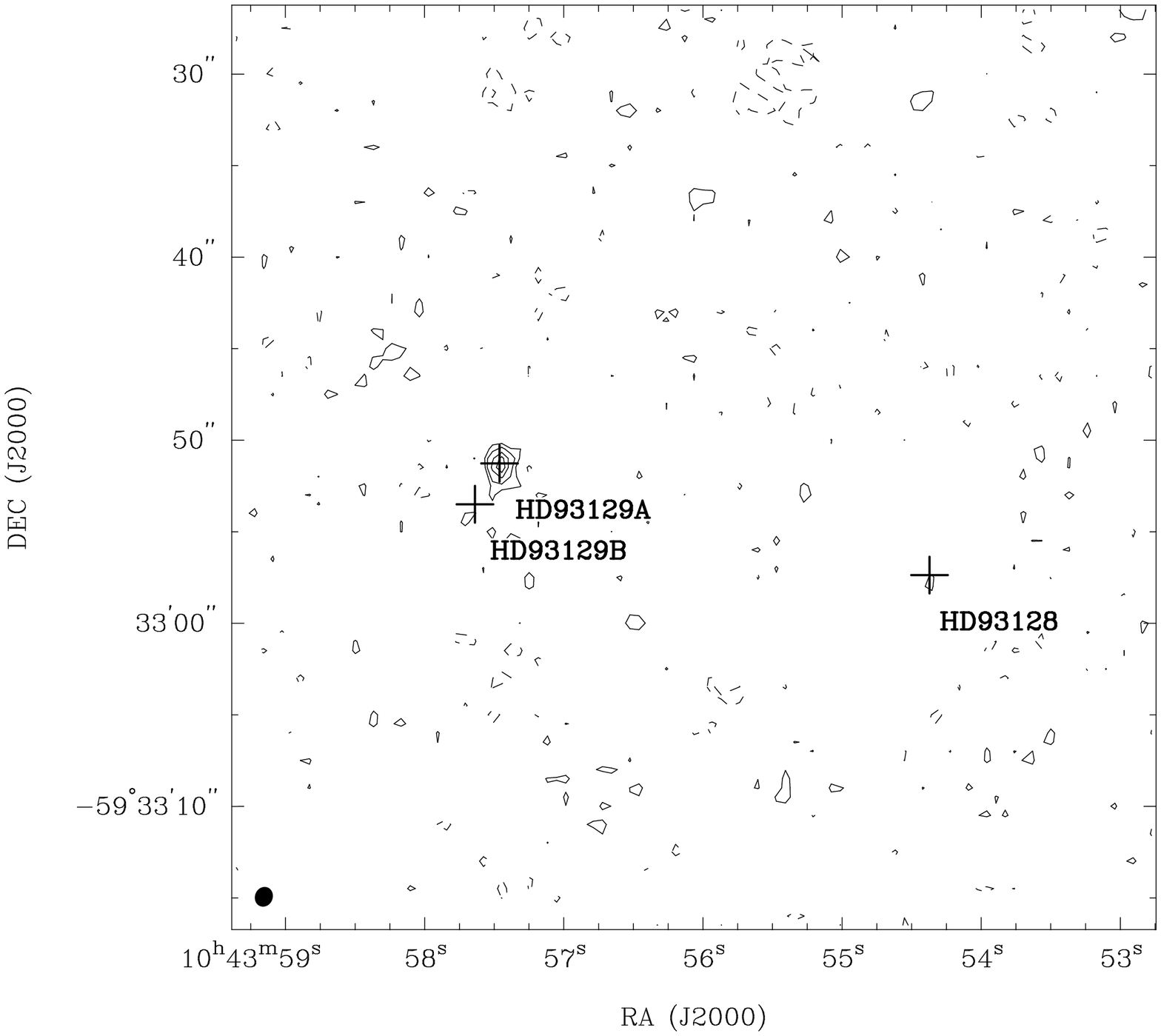,angle=-90,width=6.cm}
          \hspace*{0.4cm}
   \psfig{figure=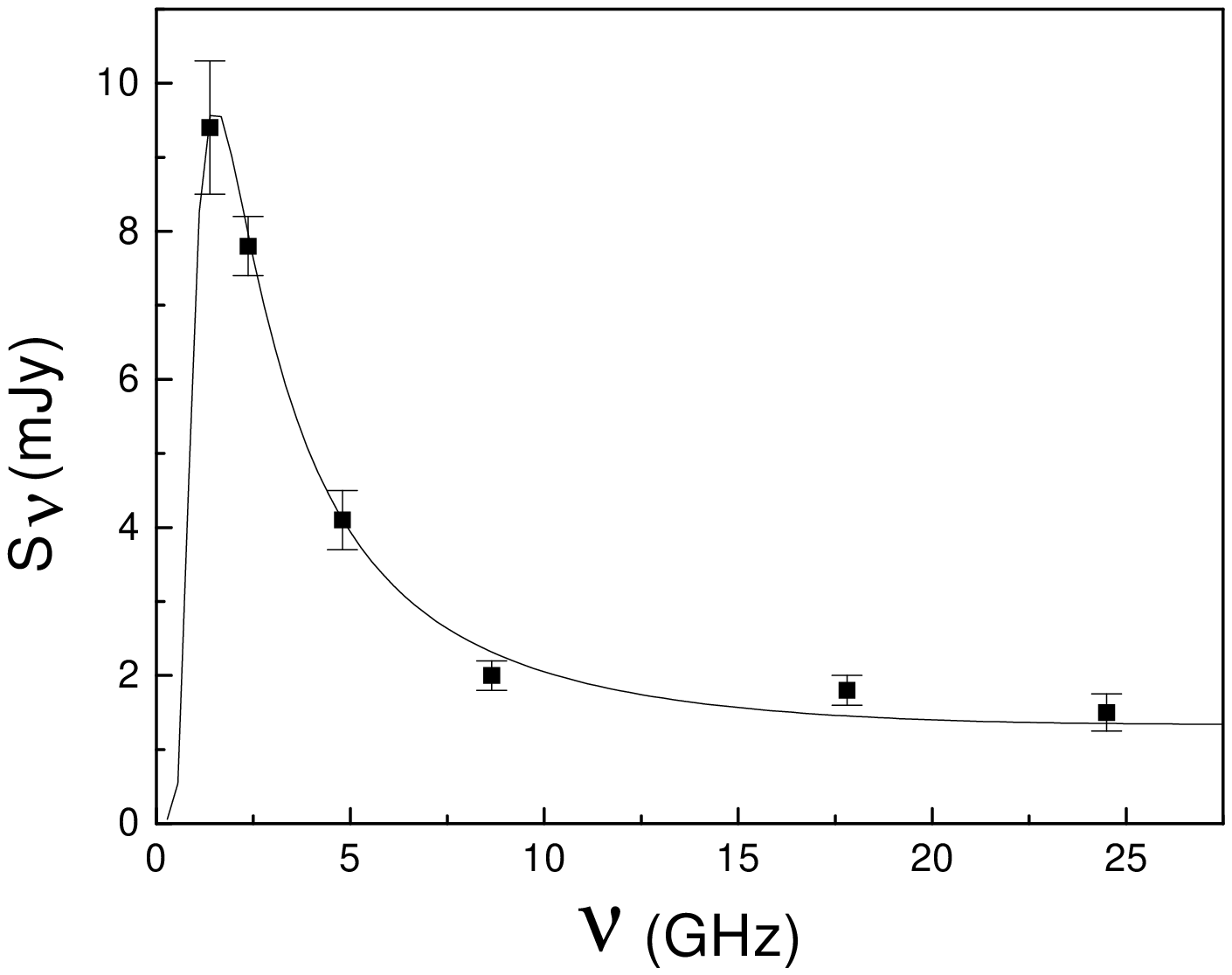,width=6.cm}
        }
\caption{{\it Left:} ATCA 3-cm radio continuum  image towards Tr
14. Optical positions of the earliest stars are marked. Contour
levels of -0.22, 0.22 (2$\sigma$), 0.44, ..., and 1.1 mJy
beam$^{-1}$. The synthesized beam ($1.1'' \times 1.0''$) 
is displayed at the bottom left corner.
{\it Right:} Radio flux density measured for HD
93129A (squares) and result of the fit (solid line).}
\end{figure}


\subsection{WR\,21a}

The star WR 21a (or Wack 2134) constitutes 
perhaps the best example to show how the
information obtained at different spectral windows aid in
assembling a complete picture of the object under study (Benaglia
et al. 2005).

WR 21a is a WN 6 star (van der Hucht 2001), $\sim$ 3 kpc from the Sun, and
its spectra showed evidence of an early-type companion (Reig
1999). It is positionally coincident with a brilliant X-ray source (1E
1022.2-5730), and with an EGRET source (3EG J1027-5817).

Spectrosocpic optical monitoring was carried out to look for a
print of the putative companion, and Niemela et al. (2006)
discovered its binary nature with a probable O companion, with a period
of weeks. A complete set of archieved X-ray data was analyzed, and
strong variability was confirmed. The available
data span over 11 yr, but the irregularly time spacing between
observations with different instruments preclude a fit with the
new period discovered.

High-angular resolution radio continuum observations were
conducted to observe the wind emission. The system was followed
with ATCA at 4.8 and 8.64 GHz. A radio source of 0.26 mJy was
detected at 4.8 GHz at the stellar position, and not at 8.64 GHz
over a r.m.s. of 0.1 mJy beam$^{-1}$. The derived spectral index
of $\alpha < 0.3$ suggests the presence of NT emission (Benaglia et al. 2005).

Low resolution HI-line radio observations were taken at IAR (HPBW
= 30'), to study the interaction between the stellar wind and the
ISM. A minimum over the position of the star was found in the HI
distribution. Unfortunately, it represents not only lack of gas (a
bubble formed by the WR+O system?), but mainly absorbed HI caused
by the strong continuum HII region source RWC 49 that lies beyond WR 21a along the same line of sight. An
HI concentration appeared on the position of the EGRET source
probability contours, with a mass $M \geq 1500$ M$_\odot$. If this cloud
is illuminated by relativistic protons accelerated at the stellar
winds, the gamma-rays produced might explain, in part, the
emission from the EGRET source.

A thoroughful HI study, to allow the
separation of emission and absorption and identify the neutral
clouds that surround the stellar system, plus gamma-ray
observations, capable of resolving the high energy stellar emission, 
are fundamental to complete the study towards this interesting system.

\begin{figure}  
\hbox{
   \psfig{figure=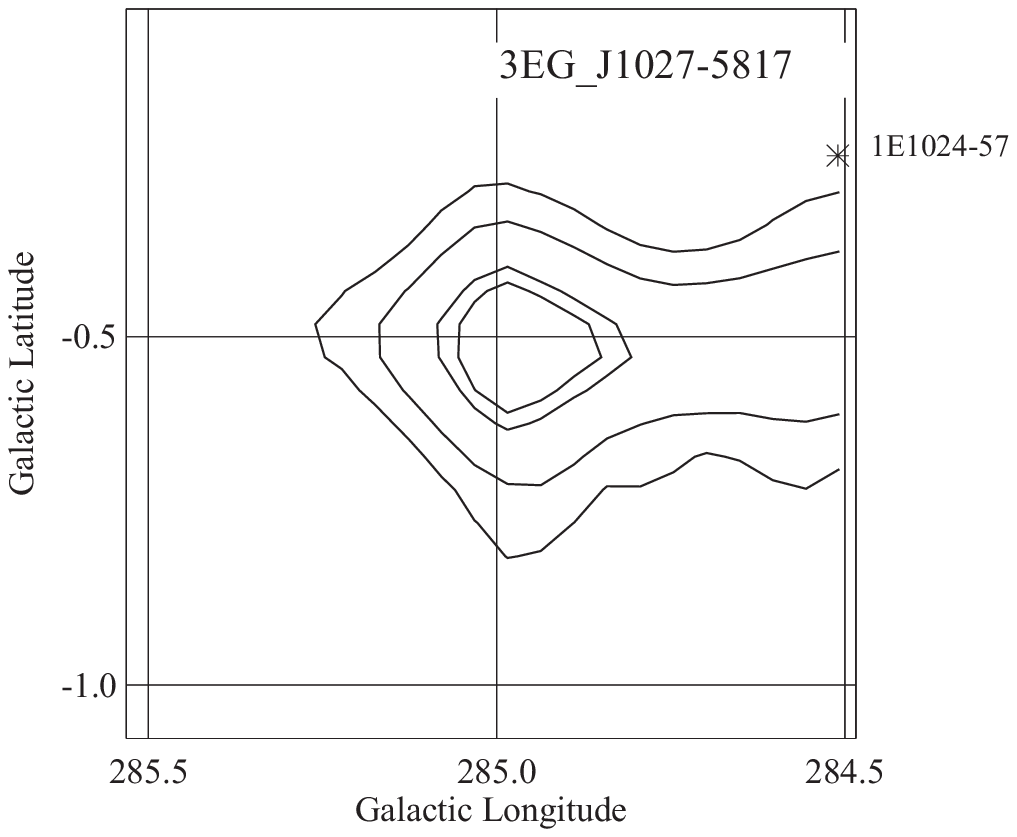,width=6.cm}
          \hspace*{0.2cm}
   \psfig{figure=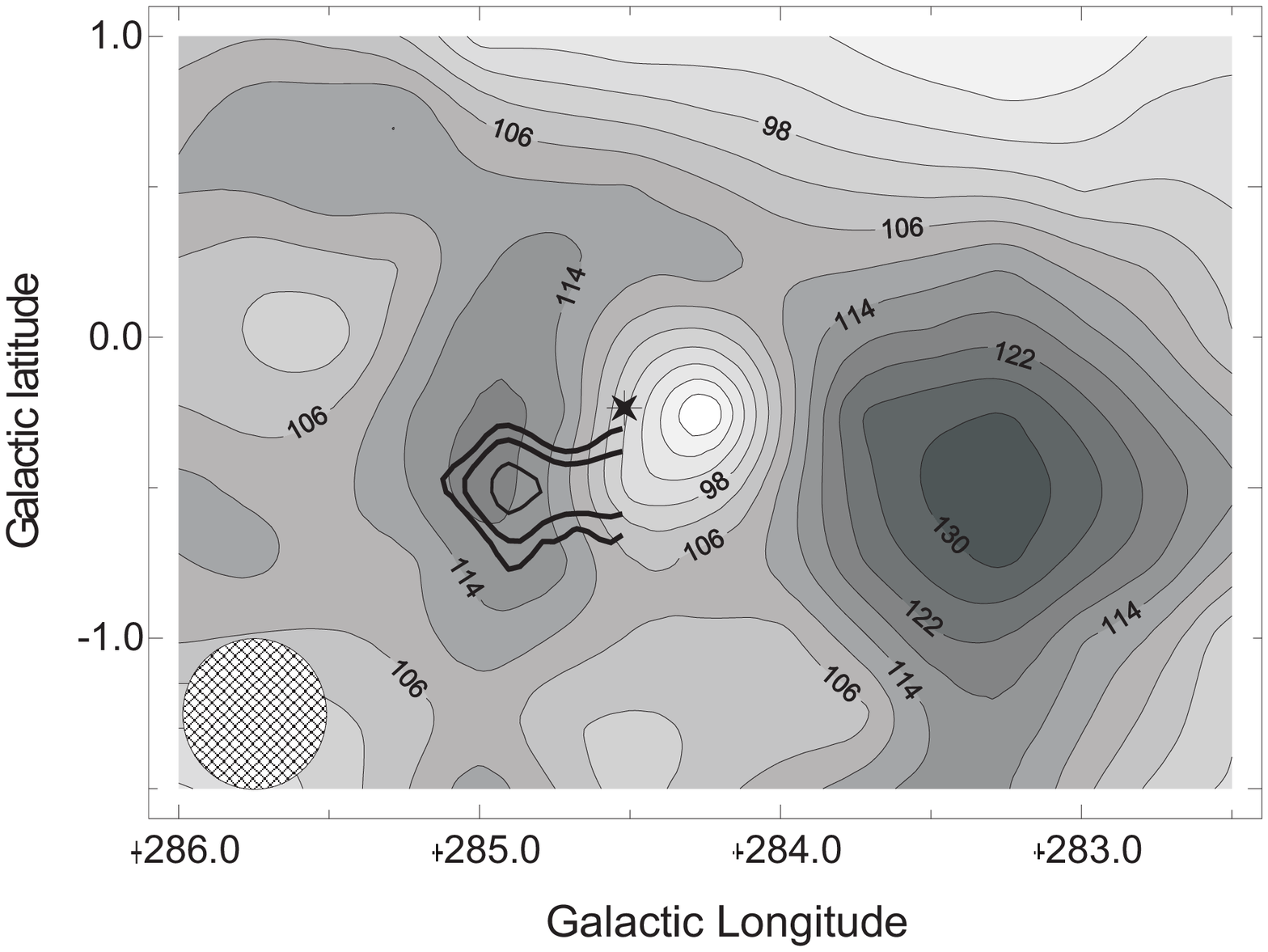,width=7.cm}
        }
\caption{{\it Left:} EGRET probability contours for 3EG
J1027-5817. Contour labels are 50\%, 68\%, 95\% and 99\%. The
X-ray source 1E 1024-57, coincident with WR~21a, is marked. {\it
Right:} Neutral-hydrogen column density integrated over velocities
from   --21 to --14 km\,s$^{-1}$. The contour levels indicate
H\,{\sc i} brightness
  temperatures in steps of 4 Kelvin. The IAR telescope beam is displayed in
  the bottom left corner. The position of WR 21a is marked with a
  black
  star; EGRET contours are superposed.}
\end{figure}

\section{Summary and perspectives}

Massive, early-type stars emit non-thermal radiation, identified
as synchrotron radiation at low energies. This fact unveils the existence of
relativistic particles and magnetic fields in the winds. The
particles are accelerated  in shocks present at the winds of
single stars, at the regions where winds of early-type stars
forming binary systems collide, and presumably at the terminal
shock where a stellar wind encounters the ISM.

The non-thermal radio emission  is detected at colliding wind
regions; the picture is not so clear for single stars. The
presence of synchrotron radiation and external photon fields suggest that high energy emission will be
produced too, detectable by means of gamma ray satellites.

Comprehensive models to reproduce  the spectral and spatial
distributions are currently under development (Reimer et al. 2005,
Pittard et al. 2005).

The number of cases in  which the emission from wind
regions could be resolved is still low: Cyg OB2 No. 5, WR 140, WR 146, WR 147.

>From the theoretical point of view, the next steps will include the
generation of 2- and 3-D hydrodynamical models up to a large wind
extent, to describe stellar winds; the application of such models
as an input to develop, in turn, more realistic models to explain
the non-thermal radio emission from single stars; the building of
self-consistent models which take into account the anti-reaction
of relativistic particles over the shock structure, etc.

Main  challenges on the observational  side are to use very long
baseline interferometry (mas resolution) to scrutinize the regions
of the stellar winds and map them, especially the ones of colliding
winds. Going even further, in time and imagination, the settlement
of an instrument of that kind to glaze at the southern stellar
wind regions -yet obscure- will be welcome by the massive stars
researchers community. These studies need to be complemented with
optical ones to investigate the structure of the stellar systems,
together with high-energy (X- and gamma-rays: Chandra, INTEGRAL,
GLAST) in order to achieve a complete picture of the stellar winds
phenomenon.

\vspace{0.4cm}

\acknowledgments I would like to thank S.M. Dougherty, who kindly
provided material included here, and G.E. Romero, for a careful reading of the manuscript. 
This work has been supported by
the Argentine agency ANPCyT through grant PICT 03-13291.

\end{document}